Viewpoint

# A Perspective on Crowdsourcing and Human-in-the-Loop Workflows in Precision Health

Peter Washington, PhD

Information and Computer Sciences, University of Hawaii at Manoa, Honolulu, HI, United States

**Corresponding Author:**
Peter Washington, PhD
Information and Computer Sciences
University of Hawaii at Manoa
1680 East-West Road
Honolulu, HI, 96822
United States
Email: pyw@hawaii.edu

## Abstract

Modern machine learning approaches have led to performant diagnostic models for a variety of health conditions. Several machine learning approaches, such as decision trees and deep neural networks, can, in principle, approximate any function. However, this power can be considered to be both a gift and a curse, as the propensity toward overfitting is magnified when the input data are heterogeneous and high dimensional and the output class is highly nonlinear. This issue can especially plague diagnostic systems that predict behavioral and psychiatric conditions that are diagnosed with subjective criteria. An emerging solution to this issue is crowdsourcing, where crowd workers are paid to annotate complex behavioral features in return for monetary compensation or a gamified experience. These labels can then be used to derive a diagnosis, either directly or by using the labels as inputs to a diagnostic machine learning model. This viewpoint describes existing work in this emerging field and discusses ongoing challenges and opportunities with crowd-powered diagnostic systems, a nascent field of study. With the correct considerations, the addition of crowdsourcing to human-in-the-loop machine learning workflows for the prediction of complex and nuanced health conditions can accelerate screening, diagnostics, and ultimately access to care.

## Introduction

Crowdsourcing, a term first coined in 2006 [1], is the use of distributed human workers to accomplish a central task. Crowdsourcing exploits the "power of the crowd" to achieve goals that are only feasible with a distributed group of humans collaborating, either explicitly or implicitly, toward a common goal. Crowdsourcing has often been applied to public health surveillance [2], such as for tracking epidemics [3,4], quantifying tobacco use [5], monitoring water quality [6], tracking misinformation [7], and understanding the black-market price of prescription opioids [8]. In the context of health care, crowdsourcing is most often used for public health, a domain that can clearly benefit from scalable and distributed assessments of health status. Although sampling bias can be an issue in epidemiological uses of crowdsourcing [9], approaches that account for these issues have performed quite robustly.

A smaller but potentially transformative effort to apply crowdsourcing to precision health rather than population health has recently emerged. In precision health contexts, the goal is to provide a diagnosis using information labeled by crowd workers. There are several variations to this basic setup. Crowdsourcing workflows for diagnostics can diverge with respect to the underlying task, worker motivation strategies, worker training, worker filtering, and privacy requirements.

Here, I describe the existing research in the relatively small and early but growing field of crowdsourcing for precision health. I then discuss ongoing challenges and corresponding opportunities that must be addressed as this field matures.

## *Existing Examples of Crowdsourcing in and Adjacent to Health Care*

There are relatively few examples of crowdsourcing in precision health. The vast successes of machine learning for health [10-15] and the human labor costs required for crowdsourcing make purely automated approaches more appealing when they are possible and feasible. However, the crowdsourcing approaches that have been tested tend to perform well for prediction tasks that are beyond the scope of current automated approaches, especially in psychiatry and the behavioral sciences.

I want to begin by highlighting successes in science, as they can often be applied to health and have started to lead to improvements in diagnostics. Framing crowdsourcing tasks as “citizen science” opportunities can be an effective incentive mechanism [16]. Oftentimes, these projects are “gamified.” Gamification refers to the incorporation of engaging elements into traditionally burdensome workflows, and in particular game-like affordances, to foster increased participation. A combination of large crowd sizes, worker training procedures, and easy identification tasks have led to previous success in the existing gamified citizen science experiments applied to precision health. For example, in a study involving nearly 100,000 crowd workers who scored images on a citizen science platform, cancer was correctly identified with an area under the receive operating characteristic of around 95% [17]. In the BioGames app, users who performed with greater than 99% accuracy in a training tutorial were invited to diagnose malaria [18,19]. It was discovered that with a large crowd size, the aggregated diagnostic accuracy of nonexpert crowd workers approached that of experts [20]. Another citizen science malaria diagnosis application, MalariaSpot, resulted in 99% accuracy in the diagnosis of malaria from blood films [21]. If the annotation task is relatively simple and nonexperts can be trained with minimal onboarding efforts, then citizen science can be an effective and affordable approach.

“Gamified” crowdsourcing for citizen science has also been successful without explicitly requiring workers to undergo a formal training process. Foldit [22-25] and EteRNA [26-31] are 2 games where players with no biology or chemistry background can explore the design space of protein and RNA folding, respectively. These are both NP-hard (ie, computationally complex) problems, and human players in aggregate have designed solutions that outcompete state-of-the-art computational approaches. These solutions have been used to solve health challenges, such as finding a gene signature for active tuberculosis, which can potentially be used in tuberculosis diagnostics [32]. Other gamified experiences have been used to build training libraries for complex classification tasks in precision psychiatry. Notably, GuessWhat is a mobile charades game played between children with autism and their parents [33,34]. While the game provides therapeutic benefits to the child with autism [35], the game simultaneously curates automatic labels of behaviors related to autism through the metadata associated with gameplay [36,37]. These automatically annotated video data have been used to develop state-of-the-art computer vision models for behaviors related to the diagnosis of autism, such as facial expression evocation [38-41], eye gaze [42], atypical prosody [43], and atypical body movements [44,45].

An alternative incentive mechanism is paid crowdsourcing. The most popular paid crowdsourcing platform, by far, is Amazon Mechanical Turk (MTurk) [46]. While paid crowdsourcing specifically for precision health is a relatively nascent field, the general study of paid crowdsourcing (particularly on MTurk) is quite mature. Studies have explored worker quality management [47], understanding crowd worker demographics [48], the generation of annotations for use in the training of machine learning models [49-53], the rights of crowd workers [54-56], and understanding crowd worker communities and economics [57-59]. Preliminary studies of paid crowdsourcing have yielded mixed success. Around 81% of images were correctly classified on MTurk in a study involving the grading of diabetic retinopathy from images, with workers failing to correctly indicate the level of severity [60]. In a separate binary labeling task for glaucomatous optic neuropathy, workers achieved sensitivity in the 80s but reached a specificity below 50% [61].

In a broader classification task of various medical conditions, workers consistently labeled the “easy” cases while struggling to correctly label and even refusing to label more complicated and nuanced tasks [62]. Clearly, there is a need for extensive innovations to the traditional paid crowdsourcing workflow to translate this methodology to precision health.

I have extensively investigated the utility of paid crowdsourcing for the diagnosis of autism from unstructured home videos, achieving relatively high diagnostic performance [63-66]. In these experiments, untrained annotators watched short videos depicting children with and without autism and answered questions about the behaviors depicted within the videos. These annotations were provided as input into previously developed machine learning models, achieving binary test performance in the 90s across performance metrics due to the reduction of the complex feature space (unstructured videos) into a low-dimensional representation (vectors of a few categorical ordinal values). This pipeline combining crowdsourcing and machine learning can possibly be extended to other diagnostic domains in psychiatry where the input feature space is complex, heterogeneous, and subjective.

## *Ongoing Challenges and Corresponding Opportunities*

Since crowdsourcing for precision health care is an emerging field of study, numerous challenges must be solved for clinical translation to develop. In the proceeding sections, I highlight several areas that are pressing for the field and for which preliminary work has been published.

### Worker Identification and Training

While traditional crowdsourcing can work with minimal to no worker training, complex annotation tasks require the identification of qualified workers. I have found that worker identification can occur through the quantification of their performance on test tasks [67,68] and training promising workers [66]. Such crowd filtration paradigms will require domain-specific procedures. There is ample room to develop new crowdsourcing systems that inherently support natural worker identification and training procedures for crowdsourcing workflows that require well-designed training processes.

### Worker Retention

Once proficient workers are identified, continually engaging and retaining these workers is critical. I have found that workers who are repeatedly encouraged by a human (or human-like chatbot) and treated as members of a broader research team tend to enjoy paid work and even ask for more tasks after the completion of the study [69]. Thus, it is possible that the guarantee of job security can lead to long-term worker retention. However, worker retention in unpaid settings that rely on intrinsic motivation will require additional innovations. For example, there exists an opportunity to explore the creation of crowd worker communities to provide a means of intrinsic motivation leading to worker retention.

### Task Assignment

Certain workers perform exceptionally well on a subset of tasks while underperforming on other assigned tasks [70,71]. There is an opportunity to develop algorithmic innovations involving the effective and optimal assignment of workers to subtasks in a dynamic manner. Reinforcement learning could be a promising approach but has yet to be explored in such scenarios.

### Privacy of Human Participants

Data in psychiatry and behavioral sciences are particularly sensitive. Ensuring that sensitive health information is handled appropriately and that workers' privacy is maintained is essential from an ethical perspective. There are 2 general families of approaches to achieving privacy in crowd-powered health care. First, the data can be modified to obscure sensitive information without removing information required for a diagnosis. I have explored privacy-preserving alterations to video data that obfuscate the identity of participants while maximizing the capacity for workers to annotate behaviors of interest [70,71]. For example, in the case of video analytics on bodily movements, the face can be tracked and blurred, or the body can be converted to a stick figure using a pose-based computer vision library. Sometimes, however, it is impossible to modify the data without severely degrading the diagnostic performance. Therefore, the second family of approaches involves carefully vetting crowd workers, training them, and onboarding them into a secure computing environment. In my previous experiences with this process [40], I discovered that crowd workers were enthusiastic about the prospect of the "job security" that is implied from the thorough vetting procedure and were, therefore, willing to complete extra privacy and security training (in our case, Research, Ethics, Compliance, and Safety training). There is ample room to expand upon these methods and to develop new paradigms and systems for crowdsourcing involving identifiable and protected health information.

### Ensuring Reliability and Reproducibility

An intrinsic challenge when incorporating human workers into precision health workflows is the variability in human responses, both within workers and between workers. I have found that while most crowd workers are inconsistent in their annotation patterns, there are workers who provide consistently sensitive and specific annotations across a wide spectrum of data points [67]. It is therefore critical to measure both internal consistency and consistency against a gold standard when recruiting crowd workers for precision health care workflows.

### Handling Financial Constraints

The crowdsourcing method with the lowest setup barriers is paid crowdsourcing. In such scenarios, financial constraints can limit the scalability of crowdsourcing workflows. One approach is to migrate from a paid system to a gamified system or another means of providing intrinsic motivation to crowd workers. However, achieving critical mass for large-scale pipelines is likely unattainable for such unpaid solutions. Paid crowd workers who consistently perform well could be recruited as full-time or long-term part-time employees for companies and organizations providing crowd-powered services. Integrating such workflows into a Food and Drug Administration (FDA)–approved process can be challenging, but it is worth exploring if it turns out that crowd-powered solutions for digital psychiatry continue to remain superior to pure-artificial intelligence (AI) approaches in the coming years.

### Translation Outside of Research Contexts

While pure machine learning approaches for precision health are beginning to translate to clinical settings through formal FDA approval procedures, the prospect of translating human-in-the-loop methods that integrate crowd workers rather than expert clinicians is daunting, especially in light of the challenges mentioned above. However, if such approaches lead to clinical-grade performance for certain conditions that are challenging to diagnose using machine learning alone, then the extra implementation and regulatory effort required to migrate these methods into production-level workflows are likely to be warranted.

## Conclusion

While machine learning for health has enabled and will continue to enable more efficient, precise, and scalable diagnostics for a variety of conditions, such models are unlikely to generalize to more difficult scenarios such as psychiatry and the behavioral sciences, which require the ability to identify complex and nuanced social human behavior. Crowd-powered human-in-the-loop workflows have the potential to mitigate some of these current limitations while still offering a high degree of automation. I invite researchers in the fields of digital phenotyping [72-76], mobile sensing [77-83], affective computing [84-90], and related subjects to consider integrating crowdsourcing and human-in-the-loop approaches into their methods when pure-AI leads to suboptimal performance.

## Acknowledgments

This project is funded by the NIH Director's New Innovator Award (DP2) from the National Institutes of Health (award DP2-EB035858).

## Conflicts of Interest

None declared.

## Abbreviations

**AI:** artificial intelligence
**FDA:** Food and Drug Administration
**MTurk:** Amazon Mechanical Turk